\documentclass[12pt]{article}  
\usepackage{graphicx}
\usepackage{txfonts}
\usepackage{epsfig}
\usepackage{bm}
\usepackage{psfrag}
\usepackage{multirow}

\newtheorem{remark}{Remark}
\newtheorem{theorem}{Theorem}

\newcommand{\erre}{{\bf r}}
\newcommand{\erredot}{\dot{\bf r}}
\newcommand{\angmom}{{\bf c}}
\newcommand{\energy}{\mathcal{E}}
\newcommand{\kepler}{\mathcal{K}}
\newcommand{\lenz}{{\bf L}}
\newcommand{\lambert}{\mathcal{L}}
\newcommand{\bX}{{\bf X}}
\newcommand{\bY}{{\bf Y}}
\newcommand{\bDelta}{\bm{\Delta}}
\newcommand{\bE}{\mathbf{E}}

\def\alphadot{\dot{\alpha}}
\def\deltadot{\dot{\delta}}
\def\da{\Delta\alpha}
\def\dd{\Delta\delta}

\def\brho{{\bm{\rho}}}
\def\vhat{\hat{\bf v}}
\def\nhat{\hat{\bf n}}
\def\erho{{\bf e}^\rho}
\def\ealpha{{\bf e}^\alpha}
\def\edelta{{\bf e}^\delta}
\def\eperp{{\bf e}^\perp}
\def\bq{{\bf q}}
\def\br{{\bf r}}

\def\rhodot{\dot{\rho}}
\def\qalpha{\bq\cdot\ealpha}
\def\qdelta{\bq\cdot\edelta}                    
\def\dq{\dot{\bf q}} 
\def\dqalpha{\dq\cdot\ealpha}
\def\dqdelta{\dq\cdot\edelta}

\def\bv{{\bf v}}
\def\bzero{{\bf 0}}
\def\bnu{\bm{\nu}}
\def\enu{{\bf e}^\nu}

\def\R{\mathbb{R}}

\def\N{\mathbb{N}}

\begin{document}
\title{On the computation of preliminary orbits for space debris with radar
  observations}
   \author{{\bf G.~F. Gronchi$^{(1)}$, L. Dimare$^{(2)}$, D. Bracali Cioci$^{(2)}$, H. Ma$^{(1)}$}\\
{\normalsize $(1)$ Dipartimento di Matematica, Universit\`a di Pisa,}\\
   \hskip 0.5cm {\normalsize Largo B. Pontecorvo 5, 56127 Pisa, Italy}\\
{\normalsize $(2)$ Space Dynamics Services s.r.l.,}\\
    \hskip 0.5cm {\normalsize Via M. Giuntini 63, 56023 Navacchio, Italy}\\
             }

   \date{}

 
\maketitle
  \begin{abstract}
We introduce a new method to perform preliminary orbit determination for
     space debris on low Earth orbits (LEO).  This method works with tracks of
     radar observations: each track is composed by $n\ge 4$ topocentric
     position vectors per pass of the satellite, taken at very short time
     intervals. We assume very accurate values for the range $\rho$, while the
     angular positions (i.e. the line of sight, given by the pointing of the
     antenna) are less accurate. We wish to correct the errors in the angular
     positions already in the computation of a preliminary orbit.
With the information contained in a pair of radar tracks, using the
     laws of the two-body dynamics, we can write 8 equations in 8
     unknowns. The unknowns are the components of the topocentric
     velocity orthogonal to the line of sight at the two mean epochs of the
     tracks, and the corrections $\bDelta$ to be applied to the angular
     positions. We take advantage of the fact that the components of
     $\bDelta$ are typically small.
We show the results of some tests, performed with simulated observations,
     and compare this algorithm with Gibbs' method 
     and the Keplerian integrals method. 
\end{abstract}


\section{Introduction}

We investigate the preliminary orbit determination problem for a satellite of
the Earth using radar observations collected by an instrument with given
technical specifications, and with a fixed observation scheduling.
Assume we collect the following data for the observed object:
\begin{equation}
\hskip 2cm 
(t_j,\rho_j,\alpha_j,\delta_j), \qquad j=1\ldots 4
\label{radtrail}
\end{equation}
where the triples $(\rho_j,\alpha_j,\delta_j)$ represent topocentric spherical 
coordinates of the object at epochs $t_j$.  Typically $\alpha_j,\delta_j$
are the values of right ascension and declination.  
We shall call {\em radar track} the set of observations in (\ref{radtrail}).

The following assumptions will be made on the data composing the tracks.
\noindent The time difference $t_{j+1}-t_j$ between consecutive observations
is $\Delta t = 10$ s. The range data $\rho_j$ are very precise: the
statistical error in the range is given by its RMS $\sigma_\rho$, which is $10$
m. On the other hand we assume that the angles $\alpha_j,\delta_j$ are not
precisely determined: their RMS $\sigma_\alpha$, $\sigma_\delta$ are supposed
to be $0.2$ degrees.
%

Given a radar track we can compute by interpolation the following data:
\begin{equation}
(\bar t, \bar\alpha, \bar\delta, \rho, \dot\rho, \ddot\rho).
\label{radinterp}
\end{equation}
Here $\bar t$, $\bar\alpha$ and $\bar\delta$ are the mean values of the epoch
and the angles, and $\rho$, $\dot\rho$, $\ddot\rho$ are the values of a
function $\rho(t)$ and its derivatives at $t=\bar t$, where $\rho(t)$ is given
by a quadratic fit with the $(t_j,\rho_j)$ data.

For low Earth orbits (LEO) these assumptions imply that the interpolated
values of $\alphadot$, $\deltadot$ are very badly accurate, to the point that
their value can be of the same order of the errors, therefore they are
practically undetermined.

By the above considerations, given a vector (\ref{radinterp}) obtained by a
radar track, and using spherical coordinates and velocities
$$
(\rho,\alpha,\delta,\dot\rho,\alphadot,\deltadot)
$$
to describe the orbit, we can consider as unknowns the quantities
$(\da,\dd,\alphadot,\deltadot)$, with
$$
\alpha = \bar\alpha + \da,\hskip 1cm
\delta = \bar\delta + \dd,
$$
where $\da, \dd$ are small deviations from the mean values
$\bar{\alpha}$, $\bar{\delta}$.

To search for the values of the unknowns we need to use additional data: we
can try to use the data of 2 radar tracks, together with a dynamical model, to
compute one or more preliminary orbits.
This is a {\em linkage problem}, see \cite{mg10}. 

In this paper we propose a new algorithm for the linkage, which takes
advantage of the smallness of $\da,\dd$, that we call {\em
  infinitesimal angles}.  We write the equations for preliminary
orbits by using the 5 algebraic integrals of Kepler's problem,
Lambert's equation for elliptic motion (see Section~\ref{s:app}) and
the projection of the equations of motion along the line of sight.

\noindent Moreover, we perform some tests to compare this method with Gibbs'
method, using only one radar track, and with the Keplerian integrals method,
which solves a linkage problem using $(\bar\alpha, \bar\delta, \rho,
\dot\rho)$ at two mean epochs (see \cite{th77}, \cite{ftmr10},
\cite{gfd11}).

The paper is organized as follows.  First we introduce some notation and
recall the basic results on Kepler's motion which are relevant for this work
(see Sections~\ref{s:eqmotion}, \ref{s:2body}, \ref{s:lambert}).  The
equations for the linkage problem, see (\ref{complete}), are presented in
Section \ref{s:linkage}, and in Sections~\ref{s:compX}, \ref{s:compDelta} we
show two different ways to compute solutions of (\ref{complete}).  In
Section~\ref{s:numexp} we present the results of some numerical tests,
including a comparison with the already known methods recalled in
Section~\ref{s:knownmethods}.  Finally, in Section~\ref{s:app}, we recall the
proof of Lambert's theorem for elliptic orbits and give a geometrical
interpretation of the results.  Moreover, we show a method to correct the
observations of a radar track so that they correspond to points in the same
plane.

\section{The equations of motion}
\label{s:eqmotion}

Let us denote by $\erho$ the unit vector corresponding to the line of sight,
and by $\bq$ the geocentric position of the observer. Then the
position of the observed body is $\br = \bq + \rho\erho$, where $\rho$ is the
range. Using the right ascension $\alpha$ and the declination $\delta$ as
coordinates we have
\[
\erho = (\cos\delta\cos\alpha, \cos\delta\sin\alpha, \sin\delta).
\]
We assume the observed body is moving according to Newton's equations
\begin{equation}
\ddot\br = -\displaystyle\frac{\mu}{|\br|^3}\br.
\label{kepeq}
\end{equation}
We introduce the moving frame $\{\erho,\vhat,\nhat\}$, depending on the epoch
$t$, where 
$\vhat=\frac{d}{ds}\erho$, regarding $\erho$ as function of the arc-length
$s$, and $\nhat=\erho\times\vhat$.
By projecting equation (\ref{kepeq}) on these vectors we obtain
\[
\left\{
\begin{array}{ll}
  \ddot{\rho} - \rho\eta^2 + \ddot{\bq}\cdot\erho &= 
  -\displaystyle\frac{\mu}{|\br|^3}(\br\cdot\erho)\cr
  2\dot{\rho}\eta + \rho\dot{\eta} + \ddot{\bq}\cdot\vhat &= 
  \ \stackrel{ }{ -\displaystyle\frac{\mu}{|\br|^3}(\br\cdot\vhat)}\cr
\kappa\eta^2\rho + \ddot{\bq}\cdot\nhat &=
\ \stackrel{ }{-\displaystyle\frac{\mu}{|\br|^3}(\br\cdot\nhat)}\cr
\end{array}
\right.
\]
where $\eta = \sqrt{\alphadot^2\cos^2\delta + \deltadot^2}$ is the proper
motion and $\kappa = \frac{d}{ds}\vhat\cdot \nhat$.  For later use we
introduce the notation
$$
  \kepler = \Bigl(\ddot\br +
  \displaystyle\frac{\mu}{|\br|^3}\br\Bigr)\cdot\erho =
  \ddot{\rho} - \rho\eta^2 + \ddot{\bq}\cdot\erho
  +\displaystyle\frac{\mu}{|\br|^3}(\br\cdot\erho).
$$

\section{The two-body integrals}
\label{s:2body}

We write below (see also \cite{gfd11}) 
the expressions of the first integrals of Kepler's
problem, i.e. the angular momentum $\angmom$, the energy $\energy$ and
the Laplace-Lenz vector $\lenz$, in the variables
$\rho,\alpha,\delta,\rhodot,\xi,\zeta$, with
\begin{equation}
\xi = \rho\alphadot\cos\delta,
\hskip 0.5cm
\zeta = \rho\deltadot.
\label{xizeta}
\end{equation}

We have
\begin{eqnarray*}
&&
\mathbf{c} = \mathbf A\xi+\mathbf B\zeta+\mathbf C,\\
&&
\energy = \frac{1}{2}|\erredot|^2 - \frac{\mu}{|\br|},\\
&&
\mu\lenz(\rho,\rhodot) = \erredot\times\angmom -
\mu\frac{\br}{|\br|} = \Bigl(\vert\erredot\vert^2
  -\frac{\mu}{|\br|}\Bigr)\br - (\erredot\cdot\br)\erredot,
\end{eqnarray*}
where
$$
\mathbf A = \br\times\ealpha,\quad
\mathbf B = \br\times\edelta,\quad 
\mathbf C = \br\times\dot\bq+
\dot\rho\,\bq\times\erho,
$$
with
$$
\ealpha = \frac{1}{\cos\delta}\frac{\partial\erho}{\partial\alpha},
\qquad
\edelta = \frac{\partial\erho}{\partial\delta},
$$
and
\begin{eqnarray*}
&&
\dot\br = \xi\ealpha + \zeta\edelta + (\dot\rho\erho + \dot\bq),\\
&&
\vert\dot\br\vert^2 = 
\xi^2 + \zeta^2 + 2\dqalpha\xi + 2\dqdelta\zeta + |\rhodot\erho + \dot\bq|^2,
\\
&&
\dot\br\cdot\br = \qalpha\xi + \qdelta\zeta + 
(\rhodot\erho + \dot\bq)\cdot\br.
\end{eqnarray*}

\noindent We introduce the notation
\[
\qquad {q}^\alpha = \qalpha,
\quad
{q}^\delta = \qdelta,
\quad
\dot{q}^\alpha = \dqalpha,
\quad
\dot{q}^\delta = \dqdelta.
\]
Note that $\xi^2 + \zeta^2 = \rho^2\eta^2$.

\section{Lambert's equation}
\label{s:lambert}

Lambert's theorem for elliptic motion gives the following relation for the
orbital elements of a body on a Keplerian orbit at epochs $t_1,t_2$:
\begin{equation}
n({t}_2-{t}_1) = \beta - \gamma - (\sin\beta -
\sin\gamma) + 2k\pi.
\label{lambert}
\end{equation}
Here $k\in\N$ is the number of revolutions in the time interval $[t_1,t_2]$,
$n=n(a)$ is the mean motion, where $a = -{\mu}/{(2\energy)}$ (the energy is
the same at the two epochs), and the angles $\beta$, $\gamma$ are defined by
\begin{equation}
\sin^2\frac{\beta}{2} =
\frac{r_1+r_2+d}{4a}, \qquad
\sin^2\frac{\gamma}{2} =
\frac{r_1+r_2-d}{4a},
\label{betagamma}
\end{equation}
and
\[
0\leq \beta-\gamma\leq 2\pi,
\]
with $r_1,r_2$ the distances from the center of force, and $d$
the length of the chord joining the two positions of the body at
epochs $t_1,t_2$.  For a fixed number of revolutions we have 4
different choices for the pairs $(\beta, \gamma)$, see
Section~\ref{s:app} and \cite{battin} for the details.

\section{Linkage}
\label{s:linkage}

We wish to link two sets of radar data (\ref{radinterp}), with mean epochs
$\bar t_i$, $i=1,2$, and compute one or more preliminary orbits.
In the following we use labels 1, 2 for the quantities introduced in
Sections~\ref{s:eqmotion}, \ref{s:2body}, \ref{s:lambert} according to the
epoch.

Let us denote by $\lambert$ the expression defining Lambert's equation. More
precisely, $\lambert=0$ is one of the possible
cases occurring in (\ref{lambert}), see Section~\ref{s:app}.
Moreover, let us define $\bv_2 =\erho_2\times\bq_2$. We consider the system
\begin{equation}
  (\angmom_1-\angmom_2,\energy_1-\energy_2,\kepler_1,\kepler_2,
  (\lenz_1-\lenz_2)\cdot\bv_2,\lambert) = {\bf 0}
\label{complete}
\end{equation}
of 8 equations in the 8 unknowns $(\bX,\bDelta)$, with
\[
\bX =
(\xi_1, \zeta_1, \xi_2, \zeta_2),
\quad \quad 
\bDelta = (\da_1, \dd_1,
\da_2, \dd_2).
\]
Note that the unknowns are divided into 2 sets so that $\bDelta$ is the
vector of infinitesimal angles.  To solve system (\ref{complete}) we
first compute $\bX$ as function of $\bDelta$ using 4 of these
equations, then we substitute $\bX(\bDelta)$ into the remaining
equations and search for solutions of the resulting nonlinear system
by applying Newton-Raphson's method. Taking advantage of the assumed
smallness of the solutions $\bDelta$, we can use $\bDelta=\bzero$ as
starting guess.

\section{Computing $\bX(\bDelta)$}
\label{s:compX}

We describe below two methods to compute $\bX$ as function of $\bDelta$ using
some of the equations of system (\ref{complete}).  One approach uses linear
equations, see Section~\ref{s:linear}, while the equations for the other are
quadratic, see Section~\ref{s:quadratic}.

\subsection{Linear equations} 
\label{s:linear}

Substituting $2\energy_1 + \rho_1{\cal K}_1 - 2\energy_2 - \rho_2{\cal
  K}_2$ in place of $\energy_1-\energy_2$ in (\ref{complete}) we
obtain an equivalent system and the equation
\begin{equation}
2\energy_1 + \rho_1{\cal K}_1 = 2\energy_2 +
\rho_2{\cal K}_2
\label{combination}
\end{equation}
is linear in the variables $\bX=(\xi_1$, $\zeta_1$, $\xi_2$, $\zeta_2)$.

\noindent Using equation (\ref{combination}) and the conservation of
the angular momentum we obtain a linear system in the variables $\bX$:
\begin{equation}
{\cal M}\bX = {\bf V}.
\label{eqforX}
\end{equation}
Here 
\[
{\cal M} = \left[
\begin{array}{cccc}
A_{11} &B_{11} &-A_{21} &-B_{21}\\
A_{12} &B_{12} &-A_{22} &-B_{22}\\
A_{13} &B_{13} &-A_{23} &-B_{23}\\
\dot{q}_1^\alpha &\dot{q}_1^\delta &-\dot{q}_2^\alpha &-\dot{q}_2^\delta \\
\end{array}
\right],
\]
where $A_{ij}, B_{ij}$ are the components of $\mathbf{A}_i, \mathbf{B}_{i}$,
and $\dot{q}_i^\alpha = \dot{\bq}_i\cdot\ealpha_i$, $\dot{q}_i^\delta =
\dot{\bq}_i\cdot\edelta_i$, for $i=1,2$.
Moreover
\[
{\bf V} = (C_{21}-C_{11},C_{22}-C_{12},C_{23}-C_{13},D_2-D_1)^T,
\]
where $C_{ij}$ are the components of $\mathbf{C}_i$ and
\[
D_i = \frac{1}{2}\Bigl(\rho_i^2\eta^2_i + |\rhodot_i\erho_i +
\dot\bq_i|^2\Bigr) -
\frac{\mu}{|\br_i|},
\]
with 
$\eta_i^2$ expressed as function of $(\da_i,\dd_i)$ by using
the equations $\kepler_i=0$, $i=1,2$, that is using relation
\[
\eta^2 =\frac{1}{\rho}\Bigl(\ddot\rho + \ddot{\bq}\cdot\erho
  +\displaystyle\frac{\mu}{|\br|^3}(\br\cdot\erho) \Bigr)
\]
at the 2 epochs $\bar{t}_1, \bar{t}_2$.

\noindent We can write $\bX$ as function of $\bDelta$ by solving system
(\ref{eqforX}).

\noindent Let us call ${\cal M}_{hj}$ the components of ${\cal M}$, and $V_h$
the components of ${\bf V}$.  The solutions of (\ref{eqforX}) are given
by
\begin{equation}
\xi_i = \frac{|{\cal M}_{2i-1}|}{|{\cal M}|},\hskip 1cm 
\zeta_i = \frac{|{\cal M}_{2i}|}{|{\cal M}|},\hskip 1cm i=1,2
\label{Xisol}
\end{equation}
where ${\cal M}_k$ has components 
\[
{\cal M}_{hj}^{(k)} = 
\left\{
\begin{array}{ll}
{\cal M}_{hj}  &\mbox{if } k\neq j\\
V_h           &\mbox{if } k= j\\
\end{array}
\right.
\]
and $|{\cal M}|$, $|{\cal M}_k|$ represent the determinants of ${\cal M}$,
${\cal M}_k$.

\subsection{Quadratic equations}
\label{s:quadratic}

The orbits at epochs $\bar{t}_1$, $\bar{t}_2$, computed with the solution
$\bX$ of system (\ref{eqforX}), do not necessarily share the same energy
$\energy$. This can produce some problems in the linear algorithm described
above, especially when solving Lambert's equation, where the right-hand sides
of (\ref{betagamma}) may become greater that $1$ during the iterations of
Newton-Raphson's method.
We can force the orbits to share the same energy by solving the first 4
equations in (\ref{complete}), that are quadratic equations in the variable
$\bX$.

\noindent By introducing the vector
\[
\bY = (\xi_1, \zeta_1, \xi_2), 
\]
we can write the conservation of
the angular momentum as the linear system
\begin{equation}
{\cal N}\bY = {\bf W}.
\label{eqforY}
\end{equation}
Here
\begin{equation}
{\cal N} = \left[
\begin{array}{ccc}
A_{11} &B_{11} &-A_{21} \\
A_{12} &B_{12} &-A_{22} \\
A_{13} &B_{13} &-A_{23} \\
\end{array}
\right]
\end{equation}
and
\[
{\bf W} = \zeta_2{\bf W}^{(1)} + {\bf W}^{(0)},
\]
where
\begin{eqnarray*}
{\bf W}^{(1)} &=& (B_{21},B_{22},B_{23})^T,\\
{\bf W}^{(0)} &=&(C_{21}-C_{11},C_{22}-C_{12},C_{23}-C_{13})^T.
\end{eqnarray*}

\noindent We solve system (\ref{eqforY}). Let us call ${\cal N}_{hj}$ the
components of ${\cal N}$ and $W_h$,$W_h^{(0)}$,$W_h^{(1)}$ the components of
${\bf W}$,${\bf W}^{(0)}$,${\bf W}^{(1)}$.  The solutions of (\ref{eqforY}) are
functions of $\zeta_2$, $\bDelta$, and are given by
\[
\tilde{\xi}_1 = \frac{|{\cal N}_1|}{|{\cal N}|},\hskip 0.5cm 
\tilde{\zeta}_1 = \frac{|{\cal N}_2|}{|{\cal N}|},\hskip 0.5cm 
\tilde{\xi}_2 = \frac{|{\cal N}_3|}{|{\cal N}|},
\]
where ${\cal N}_k$ has components 
\[
{\cal N}_{hj}^{(k)} = 
\left\{
\begin{array}{ll}
{\cal N}_{hj}  &\mbox{if } k\neq j\\
W_h           &\mbox{if } k= j\\
\end{array}
\right. .
\]
From the conservation of energy we can find $\zeta_2$ as function of
$\bDelta$.
We write
\begin{equation}
F_2\zeta_2^2 + F_1\zeta_2 + F_0 = 0,
\label{energy_cons}
\end{equation}
with
\begin{eqnarray*}
F_2 &=&\frac{1}{|{\cal N}|^2}(|{\cal N}_1^{(1)}|^2+ 
|{\cal N}_2^{(1)}|^2 - |{\cal N}_3^{(1)}|^2) - 1\\
F_1 &=&\frac{2}{|{\cal N}|^2}
(|{\cal N}_1^{(1)}||{\cal N}_1^{(0)}| + |{\cal N}_2^{(1)}||{\cal
  N}_2^{(0)}| - |{\cal N}_3^{(1)}||{\cal
  N}_3^{(0)}|) + \\
&+& \frac{2}{|{\cal N}|}(\dot{q}^\alpha_1|{\cal N}_1^{(1)}| + \dot{q}^\delta_1|{\cal
  N}_2^{(1)}| - \dot{q}^\alpha_2|{\cal N}_3^{(1)}| - \dot{q}^\delta_2|{\cal N}|)\\
F_0 &=& \frac{1}{|{\cal N}|^2}
(|{\cal N}_1^{(0)}|^2 + |{\cal N}_2^{(0)}|^2 - |{\cal N}_3^{(0)}|^2) \\
&+& \frac{2}{|{\cal N}|}(\dot{q}^\alpha_1|{\cal N}_1^{(0)}| + \dot{q}^\delta_1|{\cal
  N}_2^{(0)}| - \dot{q}^\alpha_2|{\cal N}_3^{(0)}|)\\
&+& \mathfrak{D}_1 - \mathfrak{D}_2,
\normalsize
\end{eqnarray*}
where ${\cal N}_k^{(\ell)}$, $k=1,2,3$, $\ell=0,1$, has components 
\[
{\cal N}_{hj}^{(k,\ell)} = 
\left\{
\begin{array}{ll}
{\cal N}_{hj}  &\mbox{if } k\neq j\\
W_h^{(\ell)}      &\mbox{if } k= j\\
\end{array}
\right.,
\]
and
\[
\mathfrak{D}_i = 2 D_i - \rho_i^2\eta_i^2, \qquad i=1,2.
\]
Therefore we have
\begin{eqnarray*}
\xi_1(\bDelta) &=& \tilde{\xi}_1(\zeta_2(\bDelta),\bDelta),\\
\zeta_1(\bDelta) &=& \tilde{\zeta}_1(\zeta_2(\bDelta),\bDelta),\\
\xi_2(\bDelta) &=& \tilde{\xi}_2(\zeta_2(\bDelta),\bDelta),
\end{eqnarray*}
where $\zeta_2(\bDelta)$ is a solution of (\ref{energy_cons}). 
Note that we can have up to two acceptable expressions for $\bX(\bDelta)$.

\section{Computing $\Delta$}
\label{s:compDelta}
We introduce the vector
\[
{\bf G} = (\kepler_1,\kepler_2, (\lenz_1-\lenz_2)\cdot\bv_2,{\lambert}).
\]
\begin{remark}
  To select the relevant expressions of ${\cal L}$ we need to guess
  the value of $k$ in (\ref{lambert}).  We can do this by assuming
  $\bDelta=\bzero$ and computing the possible orbits according to the
  linear or quadratic equations for $\bX(\bDelta)$.  In both cases we
  obtain two possible values for the number of revolutions $k$: with
  the linear equations we can obtain two different values of $k$ at
  the two epochs $\bar{t}_1$, $\bar{t}_2$; with the quadratic
  equations we may obtain two orbits with different $k$ at the same
  epoch, say $\bar{t}_1$, but from conservation of energy we obtain
  the same values at $\bar{t}_2$.
\end{remark}
By substituting the possible expressions of $\bX(\bDelta)$, coming
from either the linear or the quadratic equations, we obtain the
reduced system
\begin{equation}
{\cal G}(\bDelta) = {\bf G}(\bX(\bDelta),\bDelta) = {\bf 0}.
\label{reduced}
\end{equation}
Since the unknowns in $\bDelta$ are small, we can try to apply
Newton-Raphson's method with $\bDelta=\bzero$ as starting guess.
Thus we try to compute an approximation for $\bDelta$ by the iterative formula
\begin{equation}
\bDelta_{k+1} = \bDelta_k - \Bigl[\frac{\partial\cal
  G}{\partial\bDelta}(\bDelta_k)\Bigr]^{-1}{\cal G}(\bDelta_k),
\qquad \bDelta_0 = \bzero .
\label{newt_iter}
\end{equation}
Equations (\ref{newt_iter}) are linear, and are defined by (\ref{reduced}) and
by the Jacobian matrix
\[
\frac{\partial\cal G}{\partial\bDelta}({\bDelta_k}) = 
\frac{\partial{\bf G}}{\partial\bX}(\bX_k,\bDelta_k)
\frac{\partial{\bX}}{\partial\bDelta}(\bDelta_k) +
\frac{\partial{\bf G}}{\partial\bDelta}(\bX_k,\bDelta_k),
\]
with $\bX_k=\bX(\bDelta_k)$.  
\begin{remark}
Note that at each iteration the number of solutions can be doubled,
but if we impose the value of $\bDelta_{k+1}$ to be close to
$\bDelta_k$ then we can usually avoid bifurcations.
\end{remark}
The computation of the Jacobian matrix $\frac{\partial\cal
  G}{\partial\bDelta}$ is described below, enhancing the differences
between the linear and the quadratic case.

\subsection{The derivatives $\frac{\partial{\bf G}}{\partial\bX}$}

\begin{eqnarray*}
\frac{\partial \kepler_1}{\partial \bX} &=& 
-\frac{2}{\rho_1}(\xi_1,\zeta_1,0,0)\\
%
\frac{\partial \kepler_2}{\partial \bX} &=& 
-\frac{2}{\rho_2}(0,0,\xi_2,\zeta_2)
\end{eqnarray*} 
We observe that
\[
\lenz_2\cdot\bv_2 =
-\frac{1}{\mu}(\dot{\br}_2\cdot\br_2)(\dot{\br_2}\cdot\bv_2).
\]
Thus we have
\begin{eqnarray*}
\frac{\partial}{\partial \xi_1}[(\lenz_1-\lenz_2)\cdot\bv_2] &=&
\frac{2}{\mu}(\xi_1 + \dot{\bq}_1\cdot\ealpha_1)(\br_1\cdot\bv_2) \\
&&\hskip -0.3cm
-\frac{1}{\mu}[(\bq_1\cdot\ealpha_1)(\dot{\br}_1\cdot\bv_2) +
(\dot{\br}_1\cdot\br_1)(\ealpha_1\cdot\bv_2)]\\
\frac{\partial}{\partial \zeta_1}[(\lenz_1-\lenz_2)\cdot\bv_2] &=&
\frac{2}{\mu}(\zeta_1+\dot{\bq}_1\cdot\edelta_1)(\br_1\cdot\bv_2) \\
&&\hskip -0.3cm
-\frac{1}{\mu}[(\bq_1\cdot\edelta_1)(\dot{\br}_1\cdot\bv_2) 
+ (\dot{\br}_1\cdot\br_1)(\edelta_1\cdot\bv_2)]\\
\frac{\partial}{\partial \xi_2}[(\lenz_1-\lenz_2)\cdot\bv_2] &=&
 \frac{1}{\mu}[
(\bq_2\cdot\ealpha_2)(\dot{\br_2}\cdot\bv_2) +
(\dot{\br}_2\cdot\br_2)(\ealpha_2\cdot\bv_2)]\\
\frac{\partial}{\partial \zeta_2}[(\lenz_1-\lenz_2)\cdot\bv_2] &=&
 \frac{1}{\mu}[
(\bq_2\cdot\edelta_2)(\dot{\br_2}\cdot\bv_2) +
(\dot{\br}_2\cdot\br_2)(\edelta_2\cdot\bv_2)]
\end{eqnarray*} 
For Lambert's equation, the derivatives are given by
\begin{eqnarray*}
\frac{\partial \lambert}{\partial \bX} &=& \frac{\partial n}{\partial \bX}
(\bar{t}_1-\bar{t}_2)+\frac{\partial (\beta - \sin \beta)}{\partial \bX}
-\frac{\partial(\gamma-\sin \gamma)}{\partial \bX},\\
\frac{\partial n}{\partial \bX} &=& -\frac{3}{2\mu}\sqrt{-2\energy_1}
\frac{\partial (2\energy_1)}{\partial \bX},\\
\frac{\partial (\beta - \sin \beta)}{\partial \bX}&=&
(1-\cos \beta)\frac{\partial \beta}{\partial \bX}=
2\sqrt{\frac{\Gamma_+}{1-\Gamma_+}}\frac{\partial \Gamma_+}{\partial \bX},\\
\frac{\partial(\gamma-\sin \gamma)}{\partial \bX}&=&
(1-\cos \gamma)\frac{\partial \gamma}{\partial \bX}=
2\sqrt{\frac{\Gamma_-}{1-\Gamma_-}}\frac{\partial \Gamma_-}{\partial \bX},\\
\end{eqnarray*} 
with
\begin{eqnarray*} 
\Gamma_+&=&\sin^2\frac{\beta}{2}=
-\frac{r_1+r_2+d}{2\mu}\energy_1,\\
%
\Gamma_-&=&\sin^2\frac{\gamma}{2}=
-\frac{r_1+r_2-d}{2\mu}\energy_1.
\end{eqnarray*} 
In the expression for $\frac{\partial n}{\partial\bX}$ we use the energy
$\energy_1$ at epoch $\bar{t}_1$. We could as well choose $\energy_2$ at epoch
$\bar{t}_2$: this choice is arbitrary in the linear case, in fact computing
$\bX(\bDelta)$ with the linear algorithm, we generally have
$\energy_1(\xi_1(\bDelta),\zeta_1(\bDelta)) \neq
\energy_2(\xi_2(\bDelta),\zeta_2(\bDelta))$.

Since $r_1, r_2, d$ do not depend on $\bX$, we have
\begin{eqnarray*}
\frac{\partial \Gamma_+}{\partial \bX} &=& -\frac{r_1+r_2+d}{2\mu} \frac{\partial \energy_1}{\partial \bX}, \\
\frac{\partial \Gamma_-}{\partial \bX} &=& -\frac{r_1+r_2-d}{2\mu} \frac{\partial \energy_1}{\partial \bX},
\end{eqnarray*} 
with
\[
\frac{\partial \energy_1}{\partial \bX} = ((\xi_1+\bq_1\cdot\ealpha_1),
(\zeta_1+\bq_1\cdot\edelta_1),0,0).
\]

\subsection{The derivatives $\frac{\partial{\bX}}{\partial\bDelta}$}
\label{s:dXddelta}
To compute derivatives with respect to $\bDelta$ we use as intermediate
variables the unit vectors $\erho_j,\ealpha_j,\edelta_j$, $j=1,2$. 
To this aim we introduce the vector
\[
\bE=\left[\begin{array}{c}
\bE_1\cr
\bE_2\cr
\end{array}
\right],
\qquad\mathrm{where}\quad
\bE_j=\left(
\begin{array}{c}
\erho_j\\
\ealpha_j\\
\edelta_j\\
\end{array}
\right).
\]
Its derivatives with respect to $\bDelta$ are given by
\begin{eqnarray*}
\frac{\partial \bE}{\partial \bDelta} = \left[
\begin{array}{cc}
\frac{\partial \bE_1}{\partial (\alpha_1,\delta_1)} &\bzero \cr
\bzero &\frac{\partial \bE_2}{\partial (\alpha_2,\delta_2)} \cr 
\end{array}
\right],
\end{eqnarray*}
where
\[
\frac{\partial \bE_j}{\partial (\alpha_j,\delta_j)}  = 
\left[
\begin{array}{cc}
\cos\delta_j\ealpha_j  &\edelta_j\\
\eperp_j             &\bzero\\
-\sin\delta_j\ealpha_j &-\erho_j\\
\end{array}
\right]
\]
and $\eperp_j=-(\cos\alpha_j,\sin\alpha_j,0)^T$.

Moreover, we need to compute $\frac{\partial\bX}{\partial\bE}$.
We describe the different procedures for the linear and quadratic methods.

\subsubsection{The derivatives $\frac{\partial{\bX}}{\partial\bE}$, linear case}

Using (\ref{Xisol}),  we only need to compute
\begin{eqnarray*}
\frac{\partial\xi_h}{\partial\bE} &=& \frac{1}{|{\cal M}|}\frac{\partial |{\cal M}_{2h-1}| }{\partial\bE} -
\frac{|{\cal M}_{2h-1}|}{|{\cal M}|^2}\frac{\partial |{\cal M}|}{\partial\bE},\\
%
\frac{\partial\zeta_h}{\partial\bE} &=& \frac{1}{|{\cal
    M}|}\frac{\partial |{\cal M}_{2h}|}{\partial\bE} -
\frac{|{\cal M}_{2h}|}{|{\cal
    M}|^2}\frac{\partial |{\cal M}|}{\partial\bE} .
\end{eqnarray*}
We take advantage of the following relation, valid for any matrix $A$ of
order $n$ with coefficients $a_{ij}$ depending on a variable $x$:
\[
\frac{d}{dx}|A| = \sum_{h=1}^n|B_h|,
\]
where $B_h$ has coefficients $b_{ij}^{(h)}$, with
\[
b_{ij}^{(h)} = \left\{
\begin{array}{ll}
a_{ij}             &h\neq j\\
\displaystyle\frac{d}{dx}a_{ij} &h= j\\
\end{array}
\right..
\]

\subsubsection{The derivatives $\frac{\partial{\bX}}{\partial\bE}$,
  quadratic case} 
From the implicit function theorem applied to
equation (\ref{energy_cons}) we obtain
\begin{eqnarray*}
\frac{\partial\zeta_2}{\partial\bE} &=& -\left.\Bigl[\frac{1}
{(2F_2\zeta_2 + F_1)}
\Bigl(\frac{\partial F_2}{\partial\bE}\zeta_2^2 + \frac{\partial
  F_1}{\partial\bE}\zeta_2 + \frac{\partial F_0}{\partial\bE}\Bigr)\Bigr]\right|_{\zeta_2=\zeta_2^{(i)}(\bE)}.
\end{eqnarray*}
Let us define
\begin{eqnarray*}
\xi_1(\bE) &=& \tilde{\xi}_1(\zeta_2(\bE),\bE),\\
\zeta_1(\bE) &=& \tilde{\zeta}_1(\zeta_2(\bE),\bE),\\
\xi_2(\bE) &=& \tilde{\xi}_2(\zeta_2(\bE),\bE).
\end{eqnarray*}
We have
\begin{eqnarray*}
\frac{\partial\xi_1}{\partial\bE} &=& \frac{1}{|{\cal
    N}|}\Bigl(\frac{\partial |{\cal N}_{1}|}{\partial\zeta_2}\frac{\partial\zeta_2}{\partial\bE} + \frac{\partial |{\cal N}_{1}|}{\partial\bE}\Bigr)-
\frac{|{\cal N}_{1}|}{|{\cal
    N}|^2}\frac{\partial |{\cal N}|}{\partial\bE},\\
%
\frac{\partial\zeta_1}{\partial\bE} &=& \frac{1}{|{\cal
    N}|}\Bigl(\frac{\partial |{\cal N}_{2}|}{\partial\zeta_2}\frac{\partial\zeta_2}{\partial\bE} + \frac{\partial |{\cal N}_{2}|}{\partial\bE}\Bigr)-
\frac{|{\cal N}_{2}|}{|{\cal
    N}|^2}\frac{\partial |{\cal N}|}{\partial\bE},\\
%
\frac{\partial\xi_2}{\partial\bE} &=& \frac{1}{|{\cal
    N}|}\Bigl(\frac{\partial |{\cal N}_{3}|}{\partial\zeta_2}\frac{\partial\zeta_2}{\partial\bE} + \frac{\partial |{\cal N}_{3}|}{\partial\bE}\Bigr)-
\frac{|{\cal N}_{3}|}{|{\cal
    N}|^2}\frac{\partial |{\cal N}|}{\partial\bE}.\\
\end{eqnarray*}

\subsection{The derivatives  $\frac{\partial{\bf G}}{\partial\bDelta}$}

As in Section~\ref{s:dXddelta}, we compute the derivatives of $\cal G$ with
respect to $\bE$ and multiply the result by $\frac{\partial \bE}{\partial
  \bDelta}$.
We have
\[
\frac{\partial {\cal K}_j}{\partial \erho_j} = \ddot{\bq}_j +
\mu\frac{\bq_j}{|\br_j|^3}\left(1 -
  3\rho_j\frac{(\br_j\cdot\erho_j)}{|\br_j|^2}\right),
\hskip 1cm 
j=1,2
\]
and
\[
\frac{\partial {\cal K}_j}{\partial \ealpha_j} =
\frac{\partial {\cal K}_j}{\partial \edelta_j} = 0,
\hskip 1cm
\frac{\partial {\cal K}_1}{\partial \bE_2}  = \frac{\partial {\cal
    K}_2}{\partial \bE_1}  = \bzero.
\]

\begin{eqnarray*}
  \frac{\partial}{\partial \erho_1}[(\lenz_1-\lenz_2)\cdot\bv_2]  &=& 
  \frac{1}{\mu}\Bigl[(\br_1\cdot\bv_2)\left(2\rhodot_1\dot\bq_1 +
    \mu\rho_1\frac{\bq_1}{|\br_1|^3}\right) +\\
  &&\hskip -2.5cm + \left(|\dot\br_1|^2 - \frac{\mu}{|\br_1|}\right)\rho_1\bv_2 
 - (\dot\br_1\cdot\bv_2)(\rhodot_1\bq_1+\rho_1\dot\bq_1) -
 (\dot\br_1\cdot\br_1)\rhodot_1\bv_2\Bigr], \\
\frac{\partial}{\partial \ealpha_1}[(\lenz_1-\lenz_2)\cdot\bv_2]  &=& 
\frac{\xi_1}{\mu}[2(\br_1\cdot\bv_2)\dot\bq_1 - (\dot\br_1\cdot\bv_2)\bq_1 - 
(\dot\br_1\cdot\br_1)\bv_2],\\
\frac{\partial}{\partial \edelta_1}[(\lenz_1-\lenz_2)\cdot\bv_2]  &=& 
\frac{\zeta_1}{\mu}[2(\br_1\cdot\bv_2)\dot\bq_1 - (\dot\br_1\cdot\bv_2)\bq_1 - 
(\dot\br_1\cdot\br_1)\bv_2],
\end{eqnarray*}

\begin{eqnarray*}
  \frac{\partial}{\partial \erho_2}[(\lenz_1-\lenz_2)\cdot\bv_2]  &=&
-\lenz_1\times\bq_2 + \\
&&\hskip -2cm + \frac{1}{\mu}\Bigl[(\rhodot_2\bq_2+\rho_2\dot\bq_2)(\dot\br_2\cdot\bv_2) + (\dot\br_2\cdot\br_2)\bq_2\times\dot\bq_2
\Bigr],
\\
\frac{\partial}{\partial \ealpha_2}[(\lenz_1-\lenz_2)\cdot\bv_2]  &=& 
 \frac{1}{\mu}
[\xi_2(\dot\br_2\cdot\bv_2) + \zeta_2(\dot\br_2\cdot\br_2)]\bq_2,
\\
\frac{\partial}{\partial \edelta_2}[(\lenz_1-\lenz_2)\cdot\bv_2]  &=&
 \frac{1}{\mu}
[\zeta_2(\dot\br_2\cdot\bv_2) - \xi_2(\dot\br_2\cdot\br_2)]\bq_2.
\end{eqnarray*}

For Lambert's equation we have
\[
\frac{\partial \lambert}{\partial \bE} = \frac{\partial n}{\partial \bE}
(\bar{t}_1-\bar{t}_2)+\frac{\partial (\beta - \sin \beta)}{\partial \bE}
-\frac{\partial(\gamma-\sin \gamma)}{\partial \bE},
\]
with
\begin{eqnarray*}
&&\frac{\partial n}{\partial \bE_1} = -\frac{3}{2\mu}\sqrt{-2\energy_1}
\frac{\partial (2\energy_1)}{\partial \bE_1},
\hskip 1cm
\frac{\partial n}{\partial \bE_2} = {\bf 0},\\
&&\frac{\partial (\beta - \sin \beta)}{\partial \bE}=
2\sqrt{\frac{\Gamma_+}{1-\Gamma_+}}\frac{\partial \Gamma_+}{\partial \bE},\\
&&\frac{\partial(\gamma-\sin \gamma)}{\partial \bE} =
2\sqrt{\frac{\Gamma_-}{1-\Gamma_-}}\frac{\partial \Gamma_-}{\partial \bE}.
\end{eqnarray*}

\noindent Moreover
\begin{eqnarray*}
\frac{\partial \Gamma_+}{\partial \bE_1}&=& -\frac{2\energy_1}{4\mu}
(\frac{\partial r_1}{\partial \bE_1}+\frac{\partial d}{\partial \bE_1}) +
\frac{\Gamma_+}{2\energy_1}\frac{\partial (2\energy_1)}{\partial \bE_1},\\
\frac{\partial \Gamma_+}{\partial \bE_2}&=& -\frac{2\energy_1}{4\mu}
(\frac{\partial r_2}{\partial \bE_2}+\frac{\partial d}{\partial \bE_2}), \\
\frac{\partial \Gamma_-}{\partial \bE_1}&=& -\frac{2\energy_1}{4\mu}
(\frac{\partial r_1}{\partial \bE_1}-\frac{\partial d}{\partial \bE_1}) +
\frac{\Gamma_-}{2\energy_1}\frac{\partial (2\energy_1)}{\partial \bE_1},\\
\frac{\partial \Gamma_-}{\partial \bE_2}&=& -\frac{2\energy_1}{4\mu}
(\frac{\partial r_2}{\partial \bE_2}-\frac{\partial d}{\partial \bE_2}), \\
\end{eqnarray*}
with
\begin{eqnarray*}
&&\displaystyle\frac{\partial (2\energy_1)}{\partial \bE_1} = (2\dot\rho_1 \dot{\bq}_1 +
2\mu\rho_1\frac{\bq_1}{r_1^3},\ 2\xi_1\dot{\bq}_1,\ 2\zeta_1\dot{\bq}_1),\\
&&\displaystyle\frac{\partial r_1}{\partial \bE_1} = 
(\frac{\rho_1\bq_1}{r_1},{\bf 0},{\bf 0}),
\hskip 1.5cm
\displaystyle\frac{\partial r_2}{\partial \bE_2} = 
(\frac{\rho_2\bq_2}{r_2},{\bf 0},{\bf 0}),\\
&&\stackrel{}{\displaystyle\frac{\partial d}{\partial \bE_1} = 
\frac{\rho_1}{d}(\bq_1-{\bf r}_2,{\bf 0},{\bf 0}),}
\hskip 0.6cm
\displaystyle\frac{\partial d}{\partial \bE_2} =
\frac{\rho_2}{d}(\bq_2 - {\bf r}_1,{\bf 0},{\bf 0}).
\end{eqnarray*}


\section{Alternative known methods}
\label{s:knownmethods}

We recall below two already known methods that can be used in place of
the algorithm described in
Sections~\ref{s:linkage},~\ref{s:compX},~\ref{s:compDelta}, with the
available data.  An important difference is that these methods do not
provide corrections to the angles $\alpha$, $\delta$.

\subsection{Gibbs' method}
\label{s:gibbs}

From three position vectors of an observed body at the same pass we can 
compute an orbit using Gibbs' method, see \cite[Chap. 8]{herrick}.
We recall below the formulas of this method.
Given the position vectors $\erre_j$,$j=1,2,3$ at times $t_j$,
Gibbs' method gives
\[
\erredot_2=-d_1\,\erre_1 + d_2\,\erre_2 + d_3\,\erre_3,
\]
where 
\begin{eqnarray*}
d_j&=&G_j+H_j r_j^{-3},\quad j=1,2,3,\\
G_1&=&\frac{t_{32}^2}{t_{21}\,t_{32}\,t_{31}}\ , \ 
G_3=\frac{t_{21}^2}{t_{21}\,t_{32}\,t_{31}}\ , \
G_2=G_1-G_3,\\
H_1 &=& \mu\;t_{32}/12 \ , \ H_3=\mu t_{21}/12\ , \ 
H_2=H_1-H_3 .
\end{eqnarray*}
Here $t_{ij} = t_i-t_j$, $r_j=|\erre_j|$.

\subsection{Keplerian integrals}
\label{s:kepint}
From two radar tracks, we can obtain by interpolation
the values of $(\bar\alpha,\bar\delta,\rho,\dot\rho)$ 
at epochs $\bar{t}_1$, $\bar{t}_2$.
%
If we wish to determine the values of the unknowns
$\alphadot,\deltadot$, or equivalently of $\xi, \zeta$ defined by
(\ref{xizeta}),
we can use the Keplerian integrals method, see \cite{th77},
\cite{ftmr10}, \cite{gfd11}.  This method uses the equations
\[
\angmom_1 = \angmom_2, \qquad\energy_1 = \energy_2,
\]
which can be explicitly solved, giving at most two solutions.

\section{Numerical tests}
\label{s:numexp}
We have performed some numerical tests with simulated objects, without
the $J_2$ effect, but adding errors to the observations.  Here we
describe the results of our tests only for one simulated object.
We add to the data of the tracks a Gaussian error, with zero mean and standard
deviations listed in Table~\ref{t:rms}.
In particular we consider the
case where we add no error to the range $\rho$ (RMS $(1)$ in the table).
\begin{table}[h!]
\small
\begin{center}
\begin{tabular}{c|c|c|c}
RMS   &$\alpha$ &$\delta$ &$\rho$\\
\hline
$(1)$ & 0.2 & 0.2 & - \\
$(2)$ & 0.1 & 0.1 & $5\times 10^{-3}$ \\
$(3)$ & 0.2 & 0.2 & $10^{-2}$ \\
\end{tabular}
\end{center}
\caption{RMS of the errors added to the radar tracks.}
 \label{t:rms}
\end{table}
\noindent The data that we obtain by interpolation of two radar tracks are
displayed in Table~\ref{t:data} for the simulated object.  The case labeled
with RMS $(1)$ is peculiar, in fact we interpolate the available values of
$\alpha, \delta$ and we use the exact values of $\rho, \rhodot, \ddot\rho$,
that we can compute from the given orbit.
\begin{table}[h!]
\small
\begin{center}
\begin{tabular}{c|c|c|c|c}
Epoch                &Data        &RMS $(1)$   &RMS $(2)$    &RMS $(3)$\\
\hline
          &$\stackrel{}{\bar{\alpha} \ (deg)}$    &51.17      &51.20       &51.18 \\
          &$\bar{\delta} \ (deg)$    &-5.47      &-5.44       &-5.47  \\
{ 54127.155035} &$\rho \   (km)$   &1984.4     &1984.4      &1984.4 \\
{ (MJD)}   &$\rhodot \ (km/d)$ &-73313    &-73268      &-73223 \\
          &$\ddot\rho \ (km/d^2)$    &116444362   &116534247  &116676527  \\   
\hline
          &$\stackrel{}{\bar{\alpha} \ (deg)}$    &264.30     &264.28     &264.30\\
          &$\bar{\delta} \ (deg)$    &-66.77     &-66.79     &-66.77 \\
{ 54127.582118} &$\rho \   (km)$   &1893.5     & 1893.5    &1893.5  \\
{ (MJD)}   &$\rhodot \ (km/d)$ &-323582   &-323712    &-323842  \\
          &$\ddot\rho \ (km/d^2)$    &123666885  &123168829  &122613669   \\
\end{tabular}
\end{center}
\caption{Data interpolated from the radar tracks of the test object.}
\label{t:data}
\end{table}

\noindent In Table~\ref{t:20460} we show the orbits computed by the methods of
Gibbs (G), by the Keplerian integrals (KI) and by the infinitesimal angles
(InfAng), using the quadratic equations introduced in
Section~\ref{s:quadratic}.  For KI and InfAng we use the three data
sets displayed in Table~\ref{t:data}.
Note that InfAng with RMS $(1)$ is able to correct the errors in $\alpha,
\delta$ and to recover the orbital elements of the known orbit.  For RMS $(2)$
and RMS $(3)$, InfAng obtains a better value of the semimajor axis $a$, and
slightly worse values of the other orbital elements, if compared with KI.  To
be consistent, for KI with RMS $(1)$ we use the exact values of $\rho,
\rhodot$.  The results with Gibbs' method are not very good.  On the other
hand it uses only part of the information contained in the data: here we use
the three vectors $(t_j,\rho_j,\alpha_j,\delta_j)$ of the first track at
epochs $t_j$, with $j=1,2,4$.

\begin{table}[h!]
\small
\begin{center}
\begin{tabular}
{ c c | c c c | c c c}
\hline
& Known   &\multicolumn{3}{c}{OD methods} &\multicolumn{1}{|c}{RMS} \\
& orbit & \centering G   & \centering KI   & \centering InfAng     
& \\
\hline
%
%
\multirow{3}{*}{$a$} 
&         & 8267.75 & 7816.61 & 7818.10 & $(1)$\\
& 7818.10 & 8022.40 & 7815.71 & 7818.02 & $(2)$\\
&         & 8267.91 & 7814.80 & 7818.09 & $(3)$\\
\hline
\multirow{3}{*}{ $e$} 
&         & 0.005   & 0.066   & 0.066   & $(1)$\\
& 0.066   & 0.037   & 0.066   & 0.065   & $(2)$\\
&         & 0.005   & 0.066   & 0.065   & $(3)$\\
\hline
\multirow{3}{*}{ $I$} 
&         & 70.17   & 65.85   & 65.81   & $(1)$\\
& 65.81   & 68.02   & 65.84   & 65.61   & $(2)$\\
&         & 70.17   & 65.85   & 65.34   & $(3)$\\
\hline
\multirow{3}{*}{ $\Omega$} 
&         & 217.55  & 216.25  & 216.25  & $(1)$\\
& 216.25  & 216.92  & 216.25  & 216.30  & $(2)$\\
&         & 217.55  & 216.25  & 216.37  & $(3)$\\
\hline
\multirow{3}{*}{ $\omega$} 
&         & 336.76  & 357.00  & 357.16  & $(1)$\\
& 357.16  & 356.16  & 357.17  & 357.43  & $(2)$\\
&         & 335.93  & 357.19  & 357.53  & $(3)$\\
\hline
\multirow{3}{*}{ $\ell$} 
&         & 219.54  & 202.29  & 202.08  & $(1)$\\
& 202.08  & 201.46  & 202.09 & 201.72  & $(2)$\\
&         & 220.38  & 202.09  & 201.53  & $(3)$\\
\hline
\end{tabular}
\end{center}
\caption{Orbital elements at epoch $\bar{t}_1 = 54127.15$ 
  MJD computed by the tracks producing 
the three data sets of Table~\ref{t:data}. 
  Distances are expressed in km, angles in degrees.}
\label{t:20460}
\end{table}

\noindent The infinitesimal angles method with linear equations shows
some limitations, in a way that we could not compute an orbit for
reliable values of the observational errors.

\section{Conclusions}

We have introduced a new method to compute preliminary orbits of space debris
using radar observations. The comparison with already existing methods was
performed for some test cases. Large scale tests should be done, to check the
performance of the algorithm, possibly with real data.  We plan to investigate
the case which includes the $J_2$ effect in the equations: this is essential
to link radar tracks of LEO orbits after several revolutions.

\section{Acknowledgements}
This work is partially supported by the Marie Curie Initial Training
Network Stardust, FP7-PEOPLE-2012-ITN, Grant Agreement 317185.

\section{Appendix}
\label{s:app}

\subsection{Lambert's equations for elliptic motion}
\label{a:lambert}

In this section, following \cite{whittaker} and \cite{plummer}, we summarize
the steps to derive Lambert's equation for elliptic motion under a Newtonian
force and we give a geometric interpretation of the result.  Indeed we
obtain four distinct equations per number of revolutions of the observed body.
Note that, dealing with radar observations of space debris, the time between
two distinct arcs of observations usually covers several revolutions.

\begin{theorem} (Lambert, 1761) In the elliptic motion under the Newtonian
  gravitational attraction, the time $\Delta t=t_2-t_1$ spent to describe any
  arc (without multiple revolutions) from the initial position $P_1$ to the
  final position $P_2$ depends only on the semimajor axis $a$, on the
  sum $r=r_1+r_2$ of the two distances $r_1=|P_1-F|$, $r_2=|P_2-F|$ from the
  center of force $F$, and the length $d$ of the chord joining $P_1$ and
  $P_2$.  More precisely we have
\[
  n\Delta t = \beta - \gamma - (\sin\beta -
  \sin\gamma),
\]
where 
$n=n(a)$ is the mean motion, and the angles $\beta, \gamma$
are defined by
\[
\sin^2\frac{\beta}{2} =
\frac{r+d}{4a}, \qquad
\hskip 0.5cm
\sin^2\frac{\gamma}{2} =
\frac{r-d}{4a},
\]
and
\begin{equation}
0\leq \beta-\gamma\leq 2\pi.
\label{bmenog}
\end{equation}
\end{theorem}
{\em Proof.}
We can assume, without loss of generality, that the positions of the points
$P_1$, $P_2$ are defined by two values $E_1, E_2$ of the eccentric anomalies
such that $0\leq E_2-E_1\leq 2\pi$.

\noindent The difference of Kepler's equations at the two epochs gives
\[
n\Delta t= E_2-E_1-e(\sin E_2-\sin E_1),
\]
where $e$ is the orbital eccentricity. From elementary geometrical
relations we obtain
\begin{eqnarray*}
\hskip 0.5cm&& \frac{r}{a}=2\left(1-e\cos \frac{E_1+E_2}{2} \cos \frac{E_2 - E_1}{2} \right),  \\
&& \frac{d}{a}=2\sin \frac{E_2 - E_1}{2} \sqrt{1-e^2\cos^2 \frac{E_1+E_2}{2}}.
\end{eqnarray*}
It follows that
\begin{eqnarray}
\hskip 0.2cm
&& \frac{r+d}{2a}=1-\cos \left( \frac{E_2 - E_1}{2} +  \arccos \left( e\cos \frac{E_2 + E_1}{2} \right)\right), \label{eqforbeta} \\
&& \frac{r-d}{2a}=1-\cos \left(- \frac{E_2 - E_1}{2} +  \arccos \left( e\cos \frac{E_2 + E_1}{2} \right)\right).\label{eqforgamma}
\end{eqnarray}
In particular, for a real elliptical orbit to be possible the given scalar
quantities must satisfy the relations $r \geq d$ and $4a - r \geq d$.

\noindent If we define
\[
\qquad\beta_0=2\arcsin\left(\sqrt{\frac{r+d}{4a}}\right),
\hskip 0.5cm
\gamma_0=2\arcsin\left(\sqrt{\frac{r-d}{4a}}\right),
\]
then, using relation
$$
 1-\cos\theta = 2\sin^2\frac{\theta}{2}, \qquad \theta\in\R,
$$
and setting
\begin{eqnarray}
&&
\beta= \frac{E_2 - E_1}{2} +  \arccos \left( e\cos \frac{E_2 + E_1}{2} \right),  \label{beta}\\ 
&&
\gamma= - \frac{E_2 - E_1}{2} +  \arccos \left( e\cos \frac{E_2 +
    E_1}{2} \right), \label{gamma}
\end{eqnarray}
we find that the pairs
\begin{equation}
(\beta, \gamma) = 
(\beta_0, \gamma_0), \ (\beta_0, -\gamma_0), \ 
(2\pi-\beta_0, -\gamma_0), \ (2\pi-\beta_0, \gamma_0)
\label{fourpairs}
\end{equation}
satisfy equations (\ref{eqforbeta}), (\ref{eqforgamma}).

\noindent Up to addition of the same integer multiple of $2\pi$ to both
$\beta$ and $\gamma$, the pairs (\ref{fourpairs}) are the only ones fulfilling
(\ref{eqforbeta}), (\ref{eqforgamma})
and (\ref{bmenog}).
%
%

\noindent From (\ref{beta}), (\ref{gamma}) we obtain
\[
\beta - \gamma = E_2-E_1, \quad \cos\frac{\beta + \gamma}{2} = e\cos \frac{E_2 + E_1}{2},
\]
that yields
\[
n\Delta t = \beta - \gamma - \left( \sin \beta-\sin \gamma\right).
\]
In fact
\begin{eqnarray*}
\sin\beta - \sin\gamma &=& 
2\sin\frac{\beta-\gamma}{2}\cos\frac{\beta+\gamma}{2}\\
&&\hskip -1cm 
= 2e \sin\frac{E_2-E_1}{2}\cos\frac{E_2+E_1}{2} = e(\sin E_2 - \sin E_1).
\end{eqnarray*}

\rightline{$\square$}

\begin{figure}[h!]
\centering
\centerline{\includegraphics[width=9cm]{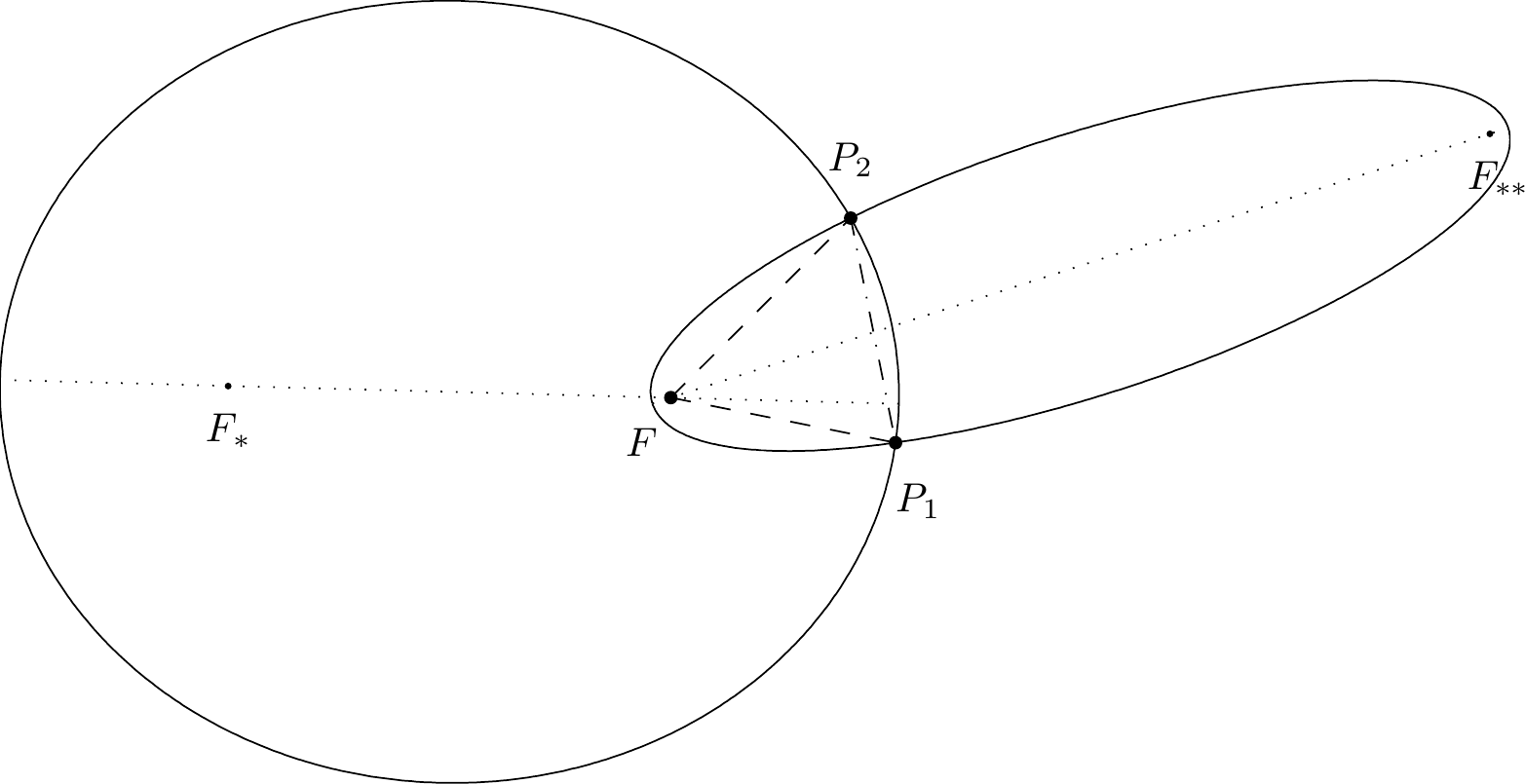}}
\caption{The four cases occurring in Lambert's theorem.}
\label{fig:4casi}
\end{figure}

The pairs $(\beta,\gamma)$ given in (\ref{fourpairs}) correspond to 4
geometrically distinct possible paths from the initial to the final position,
see Figure~\ref{fig:4casi}.  Given the points $P_1$, $P_2$ and the attracting
focus $F$, for a fixed value $a$ of the semimajor axis, we find two different
ellipses passing through $P_1$ and $P_2$. They share the attracting focus $F$,
but not the second focus ($F_*$ and $F_{**}$ in the figure). For each ellipse
we have two possible arcs from $P_1$ to $P_2$, with different orientation,
clockwise and counter-clockwise.

\noindent 
The 4 cases are discussed in \cite{p09}, \cite{plummer}, and are
distinguished on the basis of the abscissa of the intercept $Q$ of the
straight line through $P_1$ and $P_2$, on the axis passing through the foci of
one of the ellipses, measured from its center.

In \cite{plummer} the 4 cases are also distinguished using the region ${\cal
  R}$ whose border is formed by the arc and the chord joining $P_1$
and $P_2$. We use this criterion for the classification given below.

\noindent For a complete list of the equations coming from Lambert's theorem,
that need to be considered in our problem, we have to take into account the
possible occurence of multiple revolutions along the orbit.
Denoting by $n$ the mean motion, the following expressions for $\Delta t$ are
obtained:\footnote{Here the region ${\cal R}$ is defined ignoring multiple
  revolutions.}
\begin{itemize}
\item[i)] $\Delta t = T_1 + 2k\pi /n $, when the arc covers $k$
  revolutions and $\cal R$ contains neither of the foci;
\item[ii)] $\Delta t = T_2 + 2k\pi /n $, when the arc covers $k$
  revolutions, $\cal R$ contains the attracting focus $F$ but not
  the other one;
\item[iii)] $\Delta t = - T_1 + 2(k+1)\pi /n $, when the arc covers $k$
  revolutions and $\cal R$ contains both foci;
\item[iv)] $\Delta t = - T_2 + 2(k+1)\pi /n $, when the arc covers $k$
  revolutions, $\cal R$ does not contain the attracting focus $F$ but
  contains the other one;
\end{itemize}
where $T_1,T_2$ are given by
\begin{eqnarray*}
&&nT_1=\beta_0-\gamma_0 - (\sin \beta_0 - \sin \gamma_0), \nonumber \\
&&nT_2=\beta_0+\gamma_0 - (\sin \beta_0 + \sin \gamma_0).
\end{eqnarray*}
%
The four cases above can be summarized in the equation
\[
n\Delta t = \beta - \gamma - (\sin \beta - \sin \gamma) + 2k\pi,
\quad k \in \N,
\]
where the angles $\beta$, $\gamma$ are defined by 
\[
\sin^2 \frac{\beta}{2} = \frac{r+d}{4a}, 
\qquad
\sin^2 \frac{\gamma}{2} = \frac{r-d}{4a},
\] 
and
\[
0\leq \beta-\gamma\leq 2\pi.
\]

\noindent We also observe that in \cite{prussing} there is a
geometrical interpretation for the angles $\beta$, $\gamma$.

\subsection{Corrections to the observations}
\label{a:obscorr}

We describe a procedure that could be used to correct the angular
positions of a track by a pure geometrical argument.

\noindent Assume we have the geocentric position vectors $\br_j =
\rho_j\erho_j + \bq_j$, $j=1\ldots 4$, with $\bq_j$ the geocentric positions
of the observer, from the radar observations of the celestial body.  The
vectors $\br_j$ would be coplanar, if the orbit were perfectly Keplerian.  In
general this holds only approximately, due to the observational errors and to
the perturbations which should be added to Kepler's motion.
We wish to correct these position vectors and define coplanar vectors
$\br_j'$, which are slightly different from $\br_j$ and keep the
measured value $\rho_j$ of the topocentric radial distances.

In the attempt to define a good approximation of the plane of this idealized
Kepler motion, we compute the minimum of the function
\[
\bnu \mapsto Q({\bnu}) = \sum_{j=1}^4(\br_j\cdot\bnu)^2
\]
with the constraint $|\bnu|= 1$. 
We obtain the equation
\begin{equation}
\sum_{j=1}^4(\br_j\cdot\bnu)\br_j - \lambda\bnu = \bzero,
\label{Lm}
\end{equation}
with the Lagrange multiplier $\lambda\in\R$, and consider the solution
$\bnu_{min}$ of (\ref{Lm}) relative to the minimum eigenvalue $\lambda_{min}$.
We take $\enu=\bnu_{min}$ as the direction of the Kepler motion plane,
denoted by $\Pi_\nu$.

\begin{figure}[h!]
\psfragscanon
\psfrag{nu}{$\enu$}
\psfrag{r}{$\br$}\psfrag{q}{$\bq$}
\psfrag{rh}{$\brho$}\psfrag{Rrh}{${\cal R}\brho$}
\psfrag{th}{$\theta$}\psfrag{ph}{$\phi$}
\centerline{\includegraphics[width=8cm]{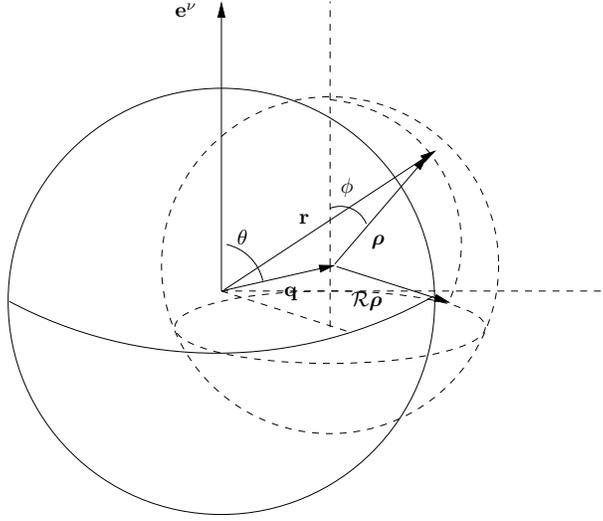}}
\psfragscanoff
\caption{Sketch of the correction of the line of sight. Here we skip the index
  $j$.}
\label{rotrho}
\end{figure}

\noindent Then, for each $j=1\ldots 4$, we rotate the vectors
$\brho_j=\rho_j\erho_j$ into a vector ${\cal R}\brho_j$ as follows
(see Figure~\ref{rotrho}).

\noindent Since we want to minimize the change in the line of sight,
i.e. the observation direction $\erho_j$, we rotate the latter around
the axis orthogonal to the plane generated by $\enu,\erho_j$ to reach
the plane $\Pi_\nu$.  In this way we draw a geodesic arc on the sphere
with radius $\rho_j$, centered at the observer position defined by
$\bq_j$.  This arc joins the position of the observed body with the
plane $\Pi_\nu$.

To describe this procedure in coordinates we introduce the angles
$\theta_j,\phi_j\in[0,\pi]$ defined by
\[
\cos\theta_j  = \frac{\enu\cdot\bq_j}{q_j},
\hskip 1cm 
\cos\phi_j = \enu\cdot\erho_j.
\]

\noindent The rotated vector ${\cal R}\brho_j$ can be expressed as the linear
combination
\[
{\cal R}\brho_j = A_j\enu + B_j\erho_j,
\]
with $B_j\ge 0$ (since we do not want to rotate the line of sight by
more than 90 degrees).

\noindent Now set the following conditions:
\begin{itemize}
\item[]\hskip 2.5cm (i)\quad $|{\cal R}\brho_j| = \rho_j$,
\item[]\hskip 2.5cm (ii)\quad $[\bq_j + {\cal R}\brho_j ]\cdot\enu = 0.$
\end{itemize}
We obtain
\begin{eqnarray*}
&&
A_j^2 + B_j^2 + 2A_jB_j\cos\phi_j = \rho_j^2,\\
&&
q_j\cos\theta_j + A_j + B_j\cos\phi_j = 0.
\end{eqnarray*}
From the second equation we obtain
\[
A_j = -q_j\cos\theta_j - B_j\cos\phi_j,
\]
that substituted into the first yields
\[
B_j = \frac{1}{\sin\phi_j}\sqrt{\rho_j^2 - q_j^2\cos^2\theta_j},
\]
so that
\[
A_j = -\left(q_j\cos\theta_j +\cot\phi_j\sqrt{\rho_j^2 -
    q_j^2\cos^2\theta_j}\right).
\]
We observe that this procedure works provided
\[
\rho_j \geq q_j\cos\theta_j,\qquad \mathrm{ for } \ \ j=1\ldots 4.
\]



\begin{thebibliography}{}

\bibitem{battin} Battin, R.~H.: {\em An introduction to the mathematics and
    methods of astrodynamics.}, AIAA Education Series, (1987)

\bibitem{ftmr10} Farnocchia, D., Tommei, G., Milani, A., Rossi, A.: {\em
    Innovative methods of correlation and orbit determination for space
    debris}, Cel. Mech. Dyn. Ast. {\bf 107}, 169--185 (2010


\bibitem{gfd11} Gronchi, G.~F., Farnocchia, D., Dimare, L.: {\em Orbit
    determination with the two-body integrals. II}, Cel. Mech. Dyn. Ast. {\bf
    110}, 257-270 (2011)

\bibitem{herrick} Herrick, S.: {\em Astrodynamics. Vol. 1.}, Van Nostrand
  Reinhold (1976)

\bibitem{mg10} Milani, A., Gronchi, G.~F.: {\em Theory of Orbit
    Determination}, Cambridge Univ. Press (2010)

\bibitem{p09} Plummer, H.~C.: {\em Lambert's theorem, some remarks on}, MNRAS
  {\bf 69}, 181 (1909)

\bibitem{plummer} Plummer, H.~C.~K.: {\em An introductory treatise on
    dynamical astronomy}, Cambridge Univ. Press (1918)

\bibitem{prussing} Prussing, J.~E.: {\em Geometrical Interpretation of the
    Angles $\alpha$ and $\beta$ in Lambert's Problem}, J. Guidance and Control
  {\bf 2/5}, 442-443 (1979)

\bibitem{th77} Taff, L.~G. and Hall, D.~L.: {\em The use of angles and angular
    rates. I - Initial orbit determination}, Celestial Mechanics {\bf 16},
  481-488 (1977)

\bibitem{whittaker} Whittaker, E.~T.: {\em A Treatise on the Analytical
    Dynamics of Particles and Rigid Bodies}, Cambridge Univ. Press, (1989)


\end{thebibliography}

\end{document}